\documentclass[openany]{nature}
\usepackage[T1]{fontenc}
\usepackage[left]{lineno}
\usepackage{amsthm,amsmath,amssymb}
\usepackage{graphicx}
\usepackage{float}
\usepackage{subfigure}
\usepackage{cases}
\usepackage{mathrsfs}
\usepackage[flushleft]{threeparttable}
\usepackage{pifont}
\usepackage{epstopdf}
\usepackage{xcolor}
\usepackage{hyperref}
\usepackage{mwe,tikz}
\usepackage{bm}
\usepackage{multirow}
\usepackage{longtable}
\usepackage{xspace}
\usepackage{rotating}
\usepackage{CJK}
\usepackage{url}
\usepackage{upgreek}
\usepackage{booktabs}
\usepackage{textcomp}
\usepackage{makecell}
\usepackage{enumitem}
\usepackage{csquotes}
\usepackage{soul}
\usepackage{tabularx}
\usepackage{overpic}
\usepackage{caption}
\usepackage{interval}
\usepackage{float}
\usepackage{threeparttable}
\usepackage{threeparttablex}
\usepackage{subfiles}
\usepackage{array}
\usepackage{makecell}
\usepackage{ulem}
\usepackage[numbers,sort&compress]{natbib}

\makeatletter
 
\def\emph#1 {\textit{ #1 } }

\let\saved@includegraphics\includegraphics
\AtBeginDocument{\let\includegraphics\saved@includegraphics}
\renewenvironment*{figure}{\@float{figure}}{\end@float}
\makeatother

\def\@fnsymbol#1{\ensuremath{\ifcase#1\or \dagger\or \ddagger\or
 \mathsection\or \mathparagraph\or \|\or **\or \dagger\dagger
 \or \ddagger\ddagger \else\@ctrerr\fi}}
\makeatother

\setlist[itemize]{leftmargin=*}

\let\oldequation\equation
\let\oldendequation\endequation
\renewenvironment{equation}{\linenomathNonumbers\oldequation}{\oldendequation\endlinenomath}

\let\oldalign\align
\let\oldendalign\endalign
\renewenvironment{align}{\linenomathNonumbers\oldalign}{\oldendalign\endlinenomath}

\let\oldgather\gather
\let\oldendgather\endgather
\renewenvironment{gather}{\linenomathNonumbers\oldgather}{\oldendgather\endlinenomath}

\newcommand{\araa}{Annu Rev Astron Astrophys}

\newcommand{\apj}{Astrophys J}
\newcommand{\apjl}{Astrophys J Lett}
\newcommand{\apjs}{Astrophys J Suppl Ser}

\newcommand{\aap}{Astron Astrophys}

\newcommand{\mnras}{Mon Not R Astron Soc}
\newcommand{\nat}{Nature}

\newcommand{\pasa}{Publ Astron Soc Aust}

\newcommand{\pasp}{Publ Astron Soc Pac}

\usepackage{newfloat}
\DeclareFloatingEnvironment[fileext=lop]{Extended Data Figure}

\title{Satellite quenching by radio jets of central galaxies in galaxy groups}

\author{Yijun Wang$^{1,2}$, Tao Wang$^{1,2}$\thanks{Corresponding authors: taowang@nju.edu.cn}, Dingyi Zhao$^{3,4}$, Yingjie Peng$^{3,4}$, Ziwen Zhang$^{5,6}$, Houjun Mo$^{7}$, Feng Yuan$^{8}$, Zhaozhou Li$^{1,2}$, Lingyu Wang$^{9,10}$, Yu Qiu$^{11}$, Yangyao~Chen$^{1,2}$, Ke Xu$^{1,2}$}

\begin{document}
%\linenumbers
\maketitle

\begin{affiliations}
\item School of Astronomy and Space Science, Nanjing University, Nanjing, Jiangsu 210023, China
\item Key Laboratory of Modern Astronomy and Astrophysics, Nanjing University, Ministry of Education, Nanjing 210023, China
\item Department of Astronomy, School of Physics, Peking University, 5 Yiheyuan Road, Beijing 100871, China
\item Kavli Institute for Astronomy and Astrophysics, Peking University, 5 Yiheyuan Road, Beijing 100871, China
\item Department of Astronomy, University of Science and Technology of China, Hefei, Anhui 230026, China
\item School of Astronomy and Space Science, University of Science and Technology of China, Hefei, 230026, China
\item Department of Astronomy, University of Massachusetts Amherst, MA 01003-9305, USA
\item Center for Astronomy and Astrophysics and Department of Physics, Fudan University, Shanghai
200438, China
\item SRON Netherlands Institute for Space Research, Landleven 12, 9747 AD Groningen, The Netherlands 
\item Kapteyn Astronomical Institute, University of Groningen, Postbus 800, 9700 AV Groningen, The Netherlands
\item College of Physics and Electronic Information Engineering, Guilin University of Technology, 319 Yanshan Street, Yanshan District, Guilin 541006, China
\end{affiliations}

\begin{abstract}
{\bf Abstract\\}
Feedback from active galactic nuclei (AGN) is now recognized as a key component of galaxy formation models. It plays a central role in regulating the growth and quenching of  galaxies in the center of groups. However, the impact of AGN feedback from central galaxies on satellite galaxies remains largely unexplored. 
Here based on the largest sample to date of radio AGNs in galaxy groups \cite{Yang2007} and a comprehensive consideration of multiple physical parameters that may influence the star formation of satellite galaxies, we demonstrate that the quiescent satellite fraction around radio AGNs is higher than that around normal galaxies. The most significant enhancement is observed around AGNs with large radio lobes. These findings demonstrate that the impact of kinetic AGN feedback beyond their host galaxies to their satellites. 
These results provide novel insights into the physical origins of some long-standing puzzles in extragalactic astronomy, including, e.g., galactic conformity and the strong small-scale clustering of quiescent galaxies.
\end{abstract}

\bigskip
\noindent{\bf keywords:} Radio galaxies; AGN feedback; Galaxy groups; Satellite galaxies; Star formation quenching; Galaxy evolution%X-ray transient; Intermediate-mass black hole; Tidal disruption event; White dwarf; Tianguan Einstein Probe

\section{Introduction}
The feedback from active galactic nuclei (AGN) has now been recognized as a central ingredient in galaxy formation models,
which may explain many observational phenomena, such as the decline of galaxy stellar mass function at massive end and luminosity function at brightest end \cite{Benson2003,Croton2006,Bower2006,Lagos2008,Schaye2015,Somerville2015}, and the scaling correlation between supermassive black holes properties and their host galaxies properties \cite{Gebhardt2000,Ferrarese2000,DiMatteo2005,Kormendy2013,Heckman2014}.
Many observational evidences and simulations suggest that AGN feedback can influence the star formation in their host galaxies \cite{McNamara2007,Fabian2012,Page2012,Harrison2017,Harrison2018,Couto2023,Zou2026}.
However, whether it can also affect the star formation in surrounding galaxies still remains poorly studied,
with galaxy groups and clusters as ideal laboratories for investigating this question. 

Mechanical feedback or jet-mode feedback from AGN are thought to be the most effective feedback mode to affect the large-scale environment  \cite{McNamara2005,Nulsen2005,Finoguenov2008,McNamara2009,Werner2010}.
So far, only several works investigate the impact of jets from central galaxies on star formation in satellite galaxies, and all of them focus on powerful radio AGNs in galaxy clusters \cite{Shabala2011,Pace2014,Shen2019}.
Therefore, it is still unclear about whether the relatively faint radio AGNs in central galaxies of galaxy groups also have an impact on the star formation in satellite galaxies, which are a more prevalent existence in our universe and are vital for helping us build a general scenario about AGN feedback on various physical scales.
In addition, due to the small sample size, previous works cannot get a consistent conclusions, with some suggesting no effect \cite{Pace2014} and others proposing quenching effect \cite{Shabala2011,Shen2019}.
Another thing worth emphasizing is that the small sample size of previous works cannot allow them to make systematic considerations for various physical properties that might bias the results, such as host halo mass, redshift, stellar mass and star formation properties of central galaxies, which had been verified to affect the star formation in satellite galaxies \cite{Weinmann2006,Kauffmann2013,MichaelaHirschmann2014,ChristianKnobel2015,HuiyuanWang2018,EnciWang2018,MartinNavarro2019}.
Therefore, there is still a lack of sufficient and reliable observational evidences about the effect of jet activities of central galaxies on star formation in satellite galaxies.

Based on the SDSS galaxy group catalog from Ref.\cite{Yang2007} and multiple radio surveys (e.g., NVSS, FIRST, and LoTSS DR2), 
we constructed the largest sample to date of radio AGNs and Fanaroff-Riley Class I/II (FR-I/FR-II) type radio sources residing in galaxy groups at $0.01 < z < 0.2$. With host halo mass, redshift, stellar mass and star formation rate of central galaxies being matched, we compare the quiescent satellite fraction around radio AGN/FR-II with that around normal galaxies to systematically investigate the intrinsic effect of jets from central galaxies on star formation in satellite galaxies. 

\section{Materials and methods}
\label{sec:DataAndMethod}

\subsection{Galaxy groups}
Galaxy groups used in this work are derived from the SDSS DR7 group catalog \cite{Yang2007} at $0.01 \leq z \leq 0.2$.
According to the brightness level in Ref.\cite{Yang2007}, the brightest galaxy in each group is considered as the `central' galaxy,
and the remaining group members are termed `satellite' galaxies.
We use brightness level rather than stellar mass level to define central and satellite galaxies because the former is more accurate than the later \cite{Weinmann2006}.
Throughout this paper, we assume a Chabrier \cite{Chabrier2003} initial mass function (IMF) and a flat cosmology with the following parameters:
$\Omega_{\rm m}=0.3$, $\Omega_{\Lambda}=0.7$, and $H_0=70\ \rm{km}\ \rm{s}^{-1}\ \rm{Mpc}^{-1}$.

\subsection{Physical parameter estimations}
Host halo mass ($M_{\rm h}$) has been verified to be the prime parameter for galaxy quenching in galaxy groups \cite{MichaelaHirschmann2014,ChristianKnobel2015,HuiyuanWang2018,EnciWang2018}.
Halo formation time is another important halo property that may affect the galaxy evolution, but it is difficult to constrain observationally, and therefore we consider only halo mass in this work. 
We adopted host halo masses derived from both Ref.\cite{Yang2007} and Ref.\cite{ZhaoDY2025}.
Ref.\cite{Yang2007} estimated $M_{\rm h}$ based on the abundance-matching (AM) method, where halo masses of galaxy groups are determined by the rank of total stellar mass of member galaxies at a given halo mass function. Ref.\cite{ZhaoDY2025} estimated $M_{\rm h}$ based on the machine learning method,
which accounts for uncertainties of halo mass estimations in AM method, such as the systematic bias between galaxy groups with star-forming centrals and quiescent centrals.
The results using $M_{\rm h}$ estimated by a new artificial neural network in Ref.\cite{Ma2025} are nearly identical to those using $M_{\rm h}$ in Ref.\cite{Yang2007}, so we do not show them in detail here.
The halo virial radius is calculated by $R_{\rm vir} = 120\ (M_{\rm h}/10^{11} M_\odot)^{1/3}$ kpc \cite{Dekel2006}.
Stellar masses ($M_\star$) and star formation rate ($SFR$) of galaxies are collected from GALEX-SDSS-WISE Legacy Catalog X2 \cite{Salim2016,Salim2018}, which are estimated by CIGALE spectral-energy-distribution-fitting code with data from ultra-violet to mid-infrared bands.
The velocity dispersion ($\sigma_\star$), 4000 {\AA} break, and emission line fluxes are from MPA-JHU DR7 release \cite{Brinchmann2004,Kauffmann2003b,Tremonti2004}.

\subsection{Radio AGN selection}
Radio AGN sample used in this work is composed of radio AGN samples from Ref.\cite{Best2012},
radio AGNs selected from VLASS survey \cite{Gordon2021},
and LoTSS DR2 radio-loud AGN sample \cite{HardcastleMJ2025}.
Ref.\cite{Best2012} used a combination of three methods to select radio AGNs from NVSS and FIRST surveys \cite{Best2012,Pace2014}.
The first method is based on 4000~{\AA} break strengths and the ratio of radio luminosity to stellar masses.
The second method is based on the ratio of radio luminosity to H$\alpha$ emission-line luminosity.
The third method is based on the standard Baldwin, Philips, \& Terlevich diagram (BPT diagram) or the emission-line diagnostic \cite{Baldwin1981,Veilleux1987,Kauffmann2003a}.
We additionally collected the VLASS 3 GHz data \cite{Gordon2021} with peak flux higher than 3 mJy beam$^{-1}$.
To correct the underestimation of flux density measurements mentioned in Ref.\cite{Gordon2021}, we followed them to scale the VLASS flux density by $1/0.87$ to compensate this underestimation. Then we used the same method in Ref.\cite{Best2012} to select radio AGNs in VLASS survey. 
Ref.\cite{Best2012} used $M_\star$ from MPA-JHU DR7 catalog based on Kroupa \cite{Kroupa2001} IMF while Ref.\cite{Salim2016,Salim2018} used Chabrier IMF. Therefore, $\log M_\star$ from Ref.\cite{Salim2016,Salim2018} is adjusted from Chabrier IMF to Kroupa IMF by applying an offset of $+0.025$ dex.
Furthermore, after correcting for the variations introduced by different IMF, Ref.\cite{Salim2016} found that their $\log M_\star$ is still 0.07 dex larger than that in Ref.\cite{Best2012}. Therefore, $\log M_\star$ applied in the first selection method\cite{Best2012} is adjusted to $\log M_\star - 0.07 + 0.025$.
The LoTSS DR2 radio-loud AGN sample\cite{HardcastleMJ2025} is selected by the relationship between radio luminosity at 144 MHz and absolute magnitude in the mid-infrared bands (WISE W2 and W3).

\subsection{Fanaroff-Riley Class II (FR-II) type radio sources selection}
FR-II type radio source sample used in this work 
is composed of FR-II type sources from Ref.\cite{Capetti2017FRII,LaoBQ2024} based on FIRST survey and from Ref.\cite{ClewsL2025} based on LOFAR survey.
We visually inspect their radio and optical images.
Then we discard the objects that have wrong classification (no radio counterpart or no extended radio morphology), 
or correspond to no or wrong optical counterpart.
We additionally exclude the sources with radio lobe size lower than 20 kpc
because some galaxy bars are misclassified as FR-II type sources by the machine learning method in Ref.\cite{LaoBQ2024}.

\subsection{Fanaroff-Riley Class I (FR-I) type radio sources selection}
FR-I type radio source sample used in this work 
is composed of FR-I type sources from Ref.\cite{Capetti2017FRI} based on FIRST survey and from Ref.\cite{ClewsL2025} based on LOFAR survey.
We visually inspect their radio and optical images.
Then we discard the objects that have wrong classification (no radio counterpart or no extended radio morphology), 
or correspond to no/wrong optical counterpart.

\subsection{Optical AGN selection}
We followed Ref.\cite{Schawinski2010} to utilize the BPT diagnostic diagrams \cite{Baldwin1981,Veilleux1987} to identify optical AGN:
(1) select sources whose [O \textsc{iii}] $\lambda$5007 luminosities are higher than $10^{40}\ \rm{erg}\ \rm{s}^{-1}$ because small number of low-ionization nuclear emission-line regions (LINERs) are above this luminosity limit \cite{Schawinski2010}; (2) further select sources with signal-to-noise ratio (S/N) of [O \textsc{iii}], H$\beta$, and H$\alpha$ higher than 3; (3) If [O \textsc{i}] $\lambda$6300 has S/N $\geq$ 3, sources beyond the extreme starburst line of Ref.\cite{Kewley2001} and the Seyfert-LINER demarcation lines from Ref.\cite{Kewley2006} in the [O \textsc{iii}]/H$\beta$ vs. [O \textsc{i}]/H$\alpha$ diagram are selected as optical AGNs; (4) If [O \textsc{i}] $\lambda$6300 has S/N $<$ 3 but [S \textsc{ii}] $\lambda$$\lambda$6717, 6731 have S/N $\geq$ 3, sources beyond the extreme starburst line of Ref.\cite{Kewley2001} and the Seyfert-LINER demarcation lines from Ref.\cite{Kewley2006} in the [O \textsc{iii}]/H$\beta$ vs. [S \textsc{ii}]/H$\alpha$ diagram are selected as optical AGNs; (5) If [O \textsc{i}] $\lambda$6300 has S/N $<$ 3 and [S \textsc{ii}] $\lambda$$\lambda$6717, 6731 have S/N $<$ 3 but [N \textsc{ii}] $\lambda$6584 has S/N $\geq$ 3, sources beyond the extreme starburst line of Ref.\cite{Kewley2001} and the Seyfert-LINER demarcation lines from Ref.\cite{Schawinski2007} in the [O \textsc{iii}]/H$\beta$ vs. [N \textsc{ii}]/H$\alpha$ diagram are selected as optical AGNs.

\subsection{Summary of samples used in this work}
In total, 23745 galaxy groups have at least one satellite galaxy with $M_\star$ above the completeness limit in Ref.\cite{Lyu2024} (hereafter ``{total-groups}'').
Here we want to stress that Ref.\cite{Lyu2024} used $M_\star$ from MPA-JHU DR7 catalog based on Kroupa IMF. After considering different IMFs and stellar mass biases (as mentioned earlier), the $M_\star$ completeness limit in Ref.\cite{Lyu2024} ($M_{\star,{\rm limit}}$) is adjusted to $\log M_{\star,{\rm limit}}+0.07-0.025$ in this work.
Among {total-groups}, we defined three target samples: (1) 2188 galaxy groups whose central galaxies hosting radio AGNs that show only compact core radio emission (hereafter ``RAGN-groups''; excluding groups whose central galaxies host FR-II or FR-I or optical AGNs); (2) 130 galaxy groups whose central galaxies hosting FR-II type sources that exhibit large and extended radio lobes (hereafter ``FR-II-groups''; excluding groups whose central galaxies host optical AGNs); (3) 544 galaxy groups whose central galaxies hosting optical AGNs (hereafter ``OptAGN-groups''; excluding groups whose central galaxies host radio AGNs or FR-II or FR-I). Their corresponding control samples are selected from 20719 galaxy groups whose central galaxies are normal galaxies (hereafter ``{normal-groups}'').
Here, normal galaxies refer to those that host neither radio AGNs, nor FR-II, nor FR-I, nor optical AGNs. In this work, we do not make detailed analysis for galaxy groups whose central galaxies host FR-I type sources because their small sample size cannot give statistically reliable conclusions and we just use FR-I sample to clean other samples.
Given that the stellar mass of central galaxies in the target and control sample has been matched in the following analysis, we do not consider the completeness of stellar mass for central galaxies in order to maximize the statistics, and only take the completeness of stellar mass into account for satellite galaxies.

\subsection{Construction of target sample and control sample} 
For each galaxy group in the target sample, its counterpart in the control sample is selected according to the following steps.
For the $n$th galaxy group in the target sample,
we label its host halo mass as $M_{{\rm h, target,}n}$,
redshift of its central galaxy as $z_{{\rm target,}n}$,
stellar mass of its central galaxy as $M_{\star,{\rm target},n}$,
and star formation rate of its central galaxy as $\textit{SFR}_{{\rm target},n}$.
From normal-groups, we select galaxy groups that satisfy the following criteria:
\begin{equation}
\begin{gathered}
\log \frac{M_{\rm h}}{M_\odot} \in [\log \frac{M_{{\rm h, target},n}}{M_\odot} - 0.10,\ \log \frac{M_{{\rm h, target},n}}{M_\odot} + 0.10], \\
{\rm and}\ z\ {\rm of\ central\ galaxy} \in [z_{{\rm target},n} - 0.04,\ z_{{\rm target},n} + 0.04], \\
{\rm and}\ \log \frac{M_\star}{M_\odot}\ {\rm of\ central\ galaxy} \in [\log \frac{M_{\star,{\rm target},n}}{M_\odot} - 0.10,\ \log \frac{M_{\star,{\rm target},n}}{M_\odot} + 0.10], \\
{\rm and}\ \log \frac{SFR}{M_\odot\ {\rm yr}^{-1}}\ {\rm of\ central\ galaxy} \in [\log \frac{SFR_{{\rm target},n}}{M_\odot\ {\rm yr}^{-1}} - 0.10,\ \log \frac{SFR_{{\rm target},n}}{M_\odot\ {\rm yr}^{-1}} + 0.10].
\end{gathered}
\end{equation}
If multiple galaxy groups in the control sample are selected out, one of them is randomly selected for inclusion in the final control sample, while the $n$th galaxy group in the target sample is added to the final target sample.
If no galaxy group in the control sample is selected out, the $n$th galaxy group in the target sample is excluded from the final target sample.
For each galaxy group in the target sample, we repeat the above steps, and construct the final target sample and final control sample.
Distributions of $M_{\rm h}$ of galaxy groups, $z$, $M_\star$, $SFR$ of central galaxies, 
and $M_\star$ of satellite galaxies are shown in Extended Data Figures \ref{efig:histYangRAGN}--\ref{efig:histZhaoBPT}.

\subsection{Identification of quiescent galaxies} 
Extended Data Figure \ref{efig1} shows $SFR$ versus $M_\star$ for all galaxies in the SDSS galaxy groups at $0.01\leq z \leq 0.2$ that lie above the adjusted $M_\star$ completeness limit from Ref.\cite{Lyu2024}.
There are two main populations in the $SFR$-$M_\star$ diagram, which are star-forming galaxies (SFGs) and quiescent galaxies (QGs).
At first, we briefly separate these two populations by the median value of specific star formation rate ($sSFR$) of the entire population.
Then we use a linear function $\log(SFR/M_\odot\ \rm{yr}^{-1}) = \alpha \log(M_\star/M_\odot) + \beta$ to fit the two main populations, respectively. Subsequently, the middle points of these two best-fit lines are taken as the reference line. Then we redefine SFGs and QGs by this reference line, i.e. SFGs are galaxies above this line in the $SFR$-$M_\star$ diagram and QGs are galaxies below this line. Then we use the above mentioned linear function to fit the redefined SFGs and QGs, and further obtained a new reference line. We iterate the procedure about redefining SFGs, QGs, and reference lines until the slope and interception of the reference line converges.
Finally, the best-fit star-forming main sequence (SFMS) is
\begin{equation}
\log \frac{\textit{SFR}}{M_\odot\ {\rm yr}^{-1}} = -4.19 + 0.39 \log \frac{M_\star}{M_\odot},
\label{eq:SFGQGdivide}
\end{equation}
which is shown as the black solid line in Extended Data Figure \ref{efig1}.  
The black dash-dotted line in Extended Data Figure \ref{efig1} is 0.9 dex below the SFMS, below which the sources are classified as QGs:
\begin{equation}
\log \frac{\textit{SFR}}{M_\odot\ {\rm yr}^{-1}} = -4.19 + 0.39 \log \frac{M_\star}{M_\odot} - 0.9.
\label{eq:SFGQGdivide}
\end{equation}

\subsection{Calculation for quiescent satellite fraction} 
We aim to measure the difference of quiescent satellite fraction ($f_{\rm q}$) between RAGN-groups/FR-II-groups/OptAGN-groups (target sample) and normal-groups (control sample) as a function of various physical parameters, such as projected distance of satellite galaxies to central galaxy and host halo mass of galaxy groups.
Firstly, to reduce the bias brought by the random selection of control sample as aforementioned, we  
repeat the selection process of target sample and control sample for 100 times.
In each parameter bin, we calculate the quiescent satellite fraction for each sample selection round, and take the median value across 100 selection rounds as the final quiescent satellite fraction.
Thus, the final quiescent satellite fraction of target sample and control sample in the $m$th parameter bin is separately defined as
\begin{gather}
f_{{\rm q,target,}m} = {\rm median}[f_{{\rm q,target,}m}^1, f_{{\rm q,target,}m}^2,...,f_{{\rm q,target,}m}^n],\ n=100, \\
f_{{\rm q,control,}m} = {\rm median}[f_{{\rm q,control,}m}^1, f_{{\rm q,control,}m}^2,...,f_{{\rm q,control,}m}^n],\ n=100,
\end{gather}
where $f_{{\rm q,target,}m}^n$ and $f_{{\rm q,control,}m}^n$ are the quiescent satellite fraction of target sample and control sample in the $m$th parameter bin in the $n$th sample selection round, respectively, which are defined as
\begin{gather}
f_{{\rm q,target,}m}^n = \frac{\sum_{i=1}^{NG_{{\rm target,}m}^n}{N_{{\rm target,}m,i,{\rm QS}}^n}}{\sum_{i=1}^{NG_{{\rm target,}m}^n}{N_{{\rm target,}m,i,{\rm TS}}^n}} \label{eq:fqtarget} \\
f_{{\rm q,control,}m}^n = \frac{\sum_{i=1}^{NG_{{\rm control,}m}^n}{N_{{\rm control,}m,i,{\rm QS}}^n}}{\sum_{i=1}^{NG_{{\rm control,}m}^n}{N_{{\rm control,}m,i,{\rm TS}}^n}}. \label{eq:fqcontrol}
\end{gather}
Here, ${NG_{{\rm target,}m}^n}$ (${NG_{{\rm control,}m}^n}$) is the number of galaxy groups in the target sample (control sample) in the $m$th parameter bin in the $n$th sample selection round, 
$N_{{\rm target,}m,i,{\rm QS}}$ ($N_{{\rm control,}m,i,{\rm QS}}$) is the number of quiescent satellite galaxies in the $i$th galaxy group of the target sample (control sample) in the $m$th parameter bin in the $n$th sample selection round,
and $N_{{\rm target,}m,i,{\rm TS}}$ ($N_{{\rm control,}m,i,{\rm TS}}$) is the number of satellite galaxies in the $i$th galaxy group of the target sample (control sample) in the $m$th parameter bin in the $n$th sample selection round.
The quiescent satellite fractions are estimated globally in parameter bin rather than averaged over the fractions of individual groups.
The $1\sigma$ uncertainties of $f_{\rm q}$ are calculated by the beta distribution intervals explained in Ref.\cite{Cameron2011}, which presents an improvement over other methods (e.g., Poisson errors) for estimating binomial confidence intervals especially for small samples.
The difference of quiescent satellite fraction between target sample and control sample is defined as
\begin{equation}
\Delta f_{\rm q} = f_{{\rm q,target}} - f_{{\rm q,control}}.
\end{equation}
Its uncertainty is estimated by error propagation. In the next analysis, we calculate $f_{\rm q}$ and $\Delta f_{\rm q}$ for satellite galaxies within 1 virial radius (1 $R_{\rm vir}$).

\section{Results}
\label{sec:results}
\begin{figure}
\centering
\includegraphics[width=\textwidth]{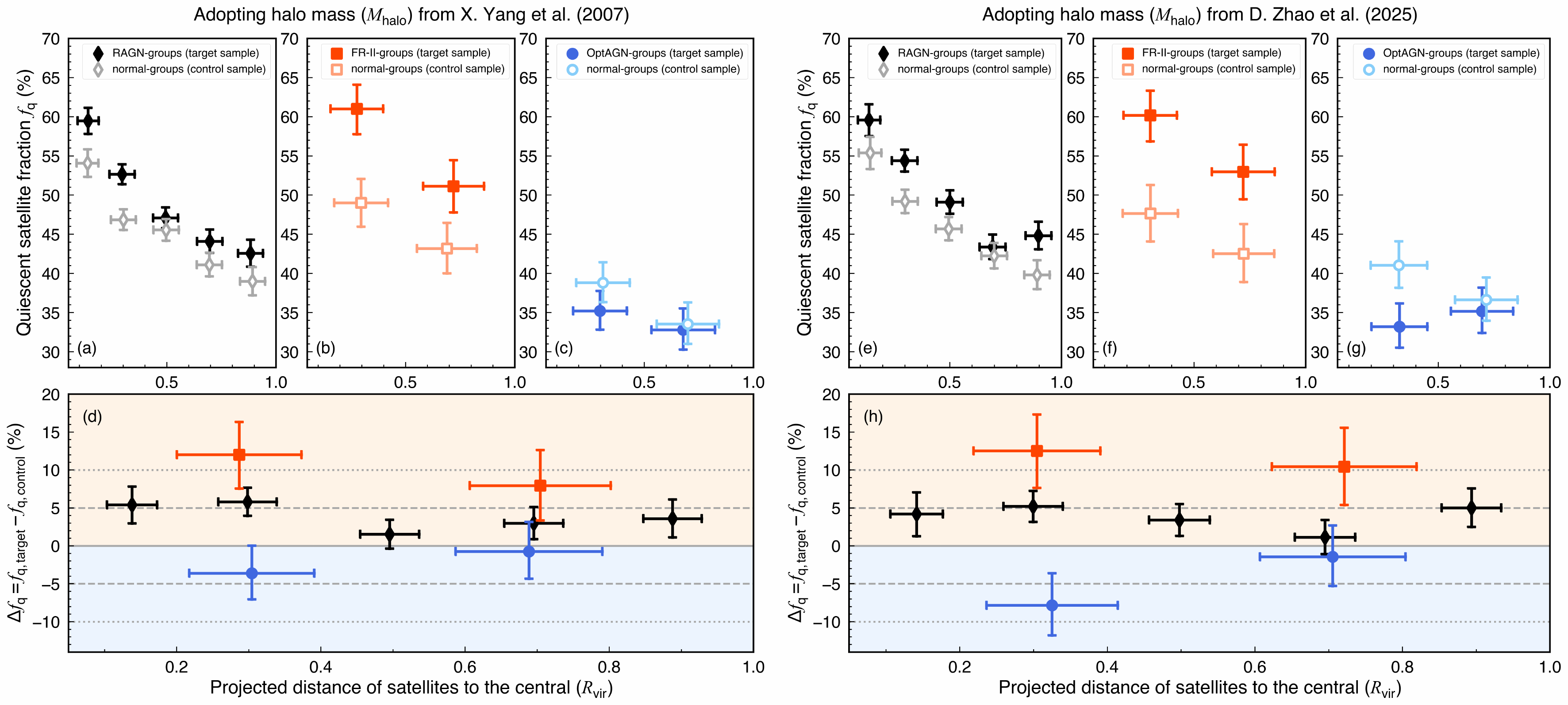}
\caption{Quiescent satellite fraction ($f_{\rm q}$) of RAGN-groups/FR-II-groups/OptAGN-groups (target sample) and normal-groups (control sample) (upper panels), and differential of $f_{\rm q}$ between target and control samples ($\Delta f_{\rm q}$) (bottom panels) as a function of projected distance of satellite galaxies to the central galaxy (in a unit of virial radius $R_{\rm vir}$). 
Left four panels and right four panels represent the results obtained with host halo mass from Ref.\cite{Yang2007} and Ref.\cite{ZhaoDY2025}, respectively.
In the upper panels, the black/red/blue symbols show RAGN-groups/FR-II-groups/OptAGN-groups (target sample), and gray/light-red/light-blue symbols show corresponding normal-groups (control sample).
In the bottom panels, the black/red/blue symbols represent $\Delta f_{\rm q}$ between RAGN-groups/FR-II-groups/OptAGN-groups and normal-groups.
The dark gray solid, dashed, and dotted lines represent $\Delta f_{\rm q} = 0\%$, $= \pm 5\%$, and $= \pm 10\%$, respectively.
The bisque region ($\Delta f_{\rm q} > 0\%$) presents that star formation of satellite galaxies around AGNs are suppressed, and the light blue region ($\Delta f_{\rm q} < 0\%$) means that star formation of satellite galaxies around AGNs are enhanced.}
\label{fig:fqdist}
\end{figure}

Figure \ref{fig:fqdist} presents $f_{\rm q}$ (upper panels) and $\Delta f_{\rm q}$ (bottom panels) as a function of projected distance of satellite galaxies to the central galaxy with $M_{\rm h}$ derived from Ref.\cite{Yang2007} (left four panels) and Ref.\cite{ZhaoDY2025} (right four panels), respectively.
Regardless of which $M_{\rm h}$ we adopt, $f_{\rm q}$ in RAGN-groups is about 5\% higher than that in normal-groups in the inner region of galaxy groups and this excess decreases slightly in the outskirts (see black points in panel (d) of Figure \ref{fig:fqdist}). The $\Delta f_{\rm q}$ between FR-II-groups and normal-groups ($\Delta f_{\rm q}\sim 10\%$) is higher than that between RAGN-groups and normal-groups.
The distributions of stellar mass of satellite galaxies between the target sample and control sample show similar shape and median value (see Extended Data Figures \ref{efig:histYangRAGN}--\ref{efig:histZhaoRAGN}), which excludes the possibility that the enhancements of $f_{\rm q}$ around radio AGNs or FR-II type sources are caused by the different distribution of stellar mass of satellite galaxies.
Radio AGNs in our sample are only composed of sources with compact core radio emission, while FR-II type sources exhibit large and extended radio lobes.
These results indicate that jet activities of AGNs in galaxy groups have a significant quenching effect on the star formation of satellite galaxies, while the quenching effect caused by large-scale radio lobes is even stronger. 
The closer the satellite galaxies are to the central radio AGNs, the more significantly their star formation is suppressed.
Prior to this, observational evidences about the effect of radio jets on the satellite galaxies in galaxy groups are lacking. 
In the outskirts of galaxy groups, the $f_{\rm q}$ around optical AGNs is similar to that around normal galaxies regardless of which $M_{\rm h}$ we adopt, while in the inner regions, $f_{\rm q}$ around optical AGNs shows no significant difference from that around normal galaxies when $M_{\rm halo}$ is from Ref.\cite{Yang2007} but is lower than that around normal galaxies when $M_{\rm halo}$ is from Ref.\cite{ZhaoDY2025}.
 Therefore, optical AGNs in central galaxies do not show a significant effect on the star formation of satellite galaxies, or only present an enhancing effect in their immediate vicinity.

\begin{figure}
\centering
\includegraphics[width=\textwidth]{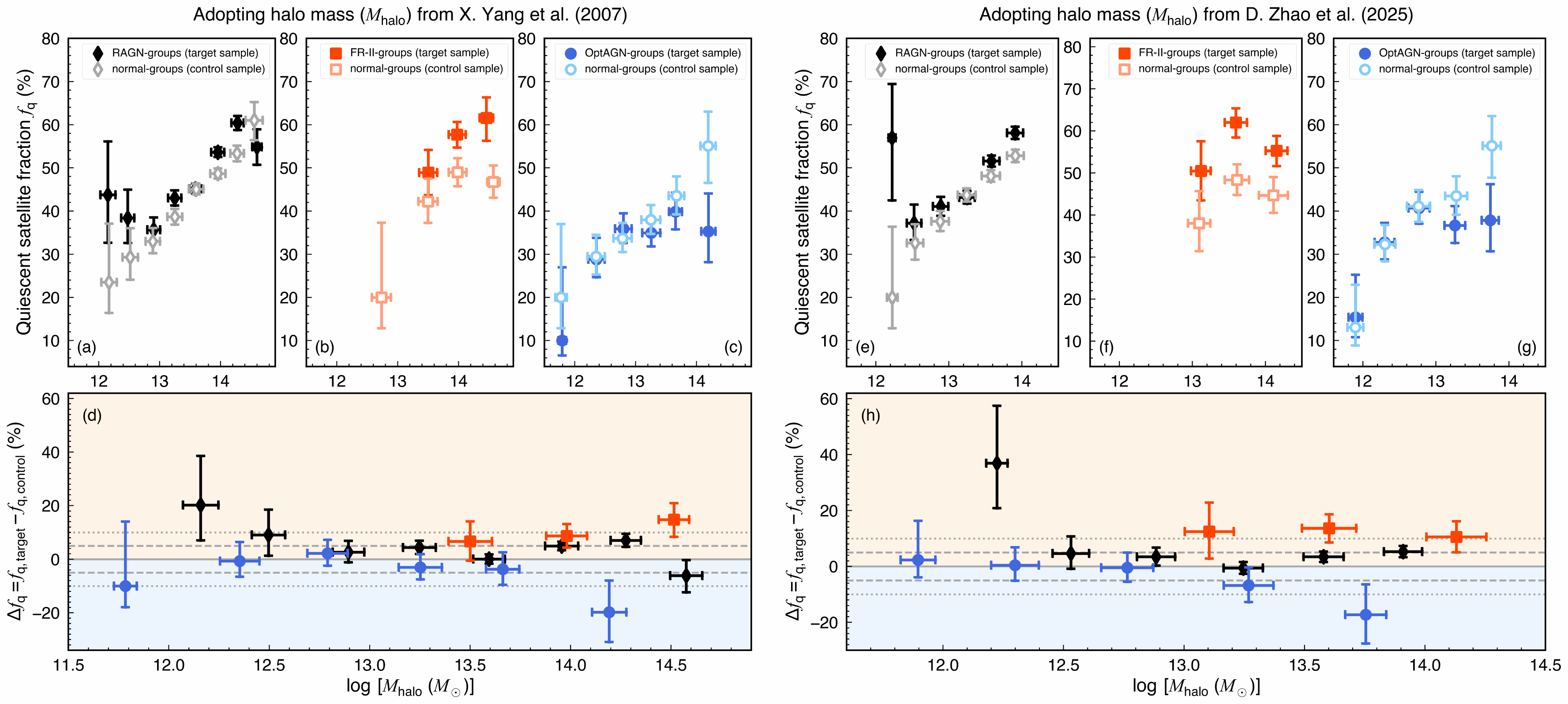}
\caption{Quiescent satellite fraction ($f_{\rm q}$) of RAGN-groups/FR-II-groups/OptAGN-groups (target sample) and normal-groups (control sample) (upper panels), and differential of $f_{\rm q}$ between target and control samples ($\Delta f_{\rm q}$) (bottom panels) as a function of host halo mass ($M_{\rm halo}$). 
Left four panels and right four panels represent the results obtained with $M_{\rm halo}$ from Ref.\cite{Yang2007} and Ref.\cite{ZhaoDY2025}, respectively.
In the upper panels, the black/red/blue symbols show RAGN-groups/FR-II-groups/OptAGN-groups (target sample), and gray/light-red/light-blue symbols show corresponding normal-groups (control sample).
In the bottom panels, the black/red/blue symbols represent $\Delta f_{\rm q}$ between RAGN-groups/FR-II-groups/OptAGN-groups and normal-groups.
The dark gray solid, dashed, and dotted lines represent $\Delta f_{\rm q} = 0\%$, $= \pm 5\%$, and $= \pm 10\%$, respectively.
The bisque region ($\Delta f_{\rm q} > 0\%$) presents that star formation of satellite galaxies around AGNs are suppressed, and the light blue region ($\Delta f_{\rm q} < 0\%$) means that star formation of satellite galaxies around AGNs are enhanced.
}
\label{fig:fqMhalo}
\end{figure}

Figure \ref{fig:fqMhalo} presents $f_{\rm q}$ (upper panels) and $\Delta f_{\rm q}$ (bottom panels) as a function of $M_{\rm h}$ with $M_{\rm h}$ derived from Ref.\cite{Yang2007} (left four panels) and Ref.\cite{ZhaoDY2025} (right four panels), respectively. Regardless of which $M_{\rm h}$ we adopt, in less massive halos ($M_{\rm halo} < 10^{12.5}\ M_\odot$), $f_{\rm q}$ around radio AGNs is significantly larger than that around normal galaxies ($\Delta f_{\rm q} \sim 20\%$--$40\%$), which indicates that radio AGNs hosted by central galaxies show a very significant quenching effect on the star formation of satellite galaxies in small halos. This quenching effect becomes weaker but still exists in massive halos ($M_{\rm halo} > 10^{12.5}\ M_\odot$). Above $M_{\rm halo} \sim 10^{13}\ M_\odot$, large-scale radio lobes of central galaxies usually have a significant quenching effect on the star formation of satellite galaxies ($\Delta f_{\rm q} \sim 5\%$--$15\%$), while below this halo mass, the lacking of enough observational data cannot allow us to draw a solid conclusion. Radio AGNs always have a quenching effect on the star formation of surrounding satellite galaxies, while this effect becomes very significant in small halos where halo quenching have not yet begun to dominate. Below $M_{\rm halo} \sim 10^{13.5}\ M_\odot$, regardless of which $M_{\rm halo}$ we adopt, optical AGNs hosted by central galaxies do not show any significant effect on the star formation of satellite galaxies. When $M_{\rm halo}$ is above $10^{{13.5}-14.0}\ M_\odot$, the star formation of satellite galaxies around optical AGNs seems to be enhanced.

\begin{figure}
\centering
\includegraphics[width=\textwidth]{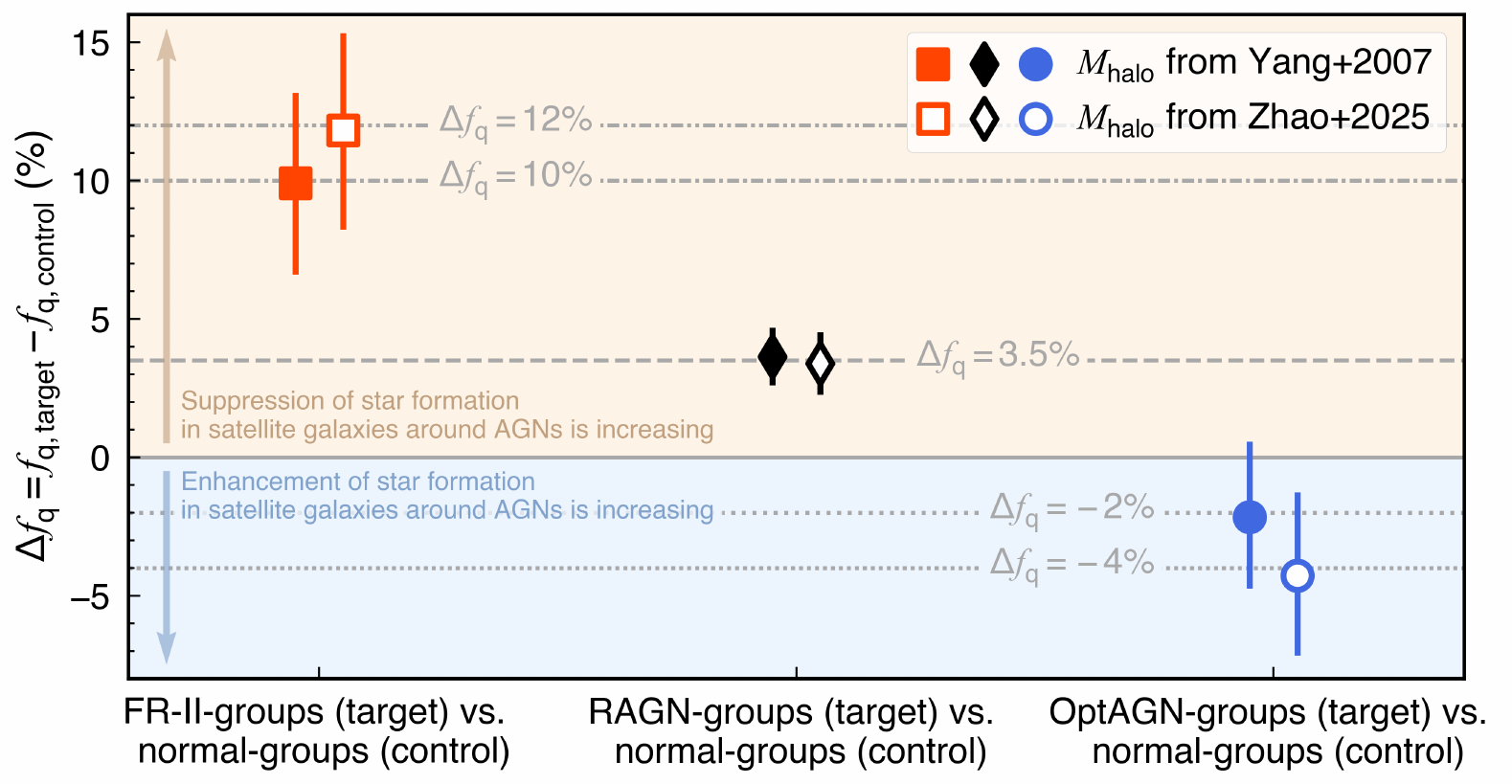}
\caption{Total differential of quiescent satellite fraction between target and control samples ($\Delta f_{\rm q}$). Red/black/blue symbols represent the comparison between FR-II-groups/RAGN-groups/OptAGN-groups and normal-groups. The solid and empty symbols show results obtained with host halo mass from Ref.\cite{Yang2007} and Ref.\cite{ZhaoDY2025}, respectively.
 The bisque region ($\Delta f_{\rm q} > 0\%$) presents that star formation of satellite galaxies around AGNs is suppressed, where the larger the $\Delta f_{\rm q}$, the more significantly the star formation of stallite galaxies around AGNs is suppressed. The light blue region ($\Delta f_{\rm q} < 0\%$) means that star formation of satellite galaxies around AGNs is enhanced, where the smaller the $\Delta f_{\rm q}$, the more significantly the star formation of satellite galaxies around AGNs is enhanced.
}
\label{fig:Totalfq}
\end{figure}

The total $\Delta f_{\rm q}$ between target sample and control sample is shown in Figure \ref{fig:Totalfq}. 
Between FR-II-groups and normal-groups, the total $\Delta f_{\rm q}$ is $9.9\%\pm 3.2\%$ for $M_{\rm halo}$ from Ref.\cite{Yang2007} and $11.8\%\pm 3.5\%$ for $M_{\rm halo}$ from Ref.\cite{ZhaoDY2025}. 
 Between RAGN-groups and normal-groups, the total $\Delta f_{\rm q}$ is $3.6\%\pm 0.9\%$ for $M_{\rm halo}$ from Ref.\cite{Yang2007} and $3.4\%\pm 1.0\%$ for $M_{\rm halo}$ from Ref.\cite{ZhaoDY2025}.
Between OptAGN-groups and normal-groups, the total $\Delta f_{\rm q}$ is $-2.2\%\pm 2.6\%$ for $M_{\rm halo}$ from Ref.\cite{Yang2007} and $-4.3\%\pm 2.8\%$ for $M_{\rm halo}$ from Ref.\cite{ZhaoDY2025}. 
Large-scale radio lobes significantly suppress the star formation of surrounding satellite galaxies, and this quenching effect becomes weaker but still exists for AGNs with small-scale radio jets, while statistically (taking the uncertainties of data points into account), optical AGNs do not show significant effect on the star formation of surrounding satellite galaxies. Our results suggest that kinetic feedback especially large-scale ones of AGNs in central galaxies have a significant quenching effect on the star formation of surrounding satellite galaxies, while under the current sample size, we find no statistically clear observational evidences for an impact of radiative feedback of AGNs on the star formation of surrounding satellite galaxies.

\section{Discussion and conclusion}
\label{sec:DiscussionAndConclusion}
Observational evidences or simulations suggest that jets, especially large-scale ones, exhibit significant feedback on their surrounding circumgalactic medium (GGM) or environments.
Powerful jets could produce strong shock to compress the GGM gas and heat them to high temperature, and the hot plasma in the large-scale radio lobes could transfer their energy to the surrounding cool CGM gas when radio lobes rise through the CGM \cite{McNamara2005,Nulsen2005,Best2007,McNamara2007,Finoguenov2008,McNamara2009,Werner2010,McNamara2012,Petter2024,HeAoyun2025}.
Cosmological magneto-hydrodynamic simulations suggest that large-scale radio lobes inject a great amount of magnetic energy into the CGM, which produce powerful magnetic pressure to suppress the gas accretion into nearby galaxies \cite{Qiu2025}.
In addition, a substantial amount of cosmic rays produced in jets or radio lobes may leak out into CGM and heat CGM gas \cite{Guo2008}.
For galaxies that have previously triggered radio jets, radio jets may be re-triggered more often or they may live longer when they are triggered \cite{Best2005}, due to the combination of  three key factors: gas supply, black hole mass, and black hole spin. In massive galaxies, a stable hot gaseous halo provides a long-term, quasi-continuous fuel source through cooling, making it easier for jet activity to be repeatedly triggered. The large black hole mass sets longer characteristic accretion timescales and higher energy budgets, allowing each active phase to persist for a longer duration and produce observable radio emission. At last, the triggering of radio jet implies that the black hole spin must be large \cite{Tchekhovskoy2010}.  These three conditions evolve on long timescales and are not erased after a single episode. The large black hole mass and spin also help enhancing the jet power, resulting in stronger heating to the CGM. These effects prevent CGM gas from cooling and falling into satellite galaxies, and finally suppress star formation in satellite galaxies.
Even if jets or radio lobes in this AGN episode disappear, their heating effect on CGM might last for a long time via turbulent mixing, even until the activities of black hole has halted or the next AGN episode begins \cite{MoHJ2002,WittorD2020,WhiteLaurel2025,HeAoyun2025}. 

The quenching effect of jets from central galaxies on satellite galaxies also provide novel insights into the physical origin of some longstanding puzzles in extragalactic astronomy, such as ``galactic conformity'' phenomenon and the strong small-scale clustering of quiescent galaxies at $r_{\rm p} < 1-2 h^{-1}$ Mpc.
The star formation properties of satellite galaxies have been verified to correlate with the star formation properties of central galaxies, a phenomenon known as ``galactic conformity'' \cite{Weinmann2006,Prescott2011,Kauffmann2013,Hartley2015,Wang2015,Berti2017,Otter2020,McConachie2025}. This phenomenon creates new challenges to our galaxy formation models, but its origin still remain unclear.
Although AGN feedback has been proposed as one possible reason \cite{Weinmann2006}, direct observational evidences are still lacking.
Our findings that jets from central galaxies in galaxy groups could quench the star formation in satellite galaxies may provide important clues about the origins of the galactic conformity, because the jet is expected to be also able to suppress star formation of the central galaxy. Moreover, quiescent galaxies have higher probabilities to trigger jets, particularly those with strong lobes. Together, this could explain the presence of a higher quiescent satellite fraction around quiescent centrals, as well as the strong small-scale clustering for quiescent galaxies. Recently, JWST observations have revealed a number of cases at $z \sim 3-4$ where quiescent galaxies may be significantly affected, if not directly quenched, by neighbouring massive galaxies hosting powerful AGNs\cite{PerezGonzalez2025,Wang2026}. It suggests that the impact of AGNs on its satellites and even neighbouring centrals should be considered as a major mechanism in quenching galaxies over cosmic time, which has been unfortunately missed in previous galaxy formation models and simulations. 

\clearpage

%%%%%%%%%%figures and tables in the Main part begin here%%%%%%%%

\captionsetup[table]{name={\bf Table}}
\captionsetup[figure]{name={\bf Fig.}}

\clearpage

\clearpage

\clearpage

\begin{addendum}

\item[Conflict Interests] The authors declare that they have no conflict of interest.

\item[Acknowledgments]  This work was supported by National Natural Science Foundation of China (Grant No. 12403019, 12525302, and 12141301), Jiangsu Natural Science Foundation (Grant No. BK20241188), Basic Research Program of Jiangsu (Grant No. BK20250001), National Key R\&D Program of China (Grant No. 2023YFA1605600), the Fundamental Research Funds for the Central Universities with Grant no.KG202502, and the China Manned Space Program with grant no. CMS-CSST-2025-A04. Yu Qiu acknowledges supports by National Natural Science Foundation of China (Grant No. 12573011).

\item[Author Contributions] Yijun Wang and Tao Wang initiated the study. Yijun Wang selected the sample, made data analysis, and authored the majority of the text under the supervision of Tao Wang. Dingyi Zhao, Yingjie Peng, Ziwen Zhang, Houjun Mo, Feng Yuan, Zhaozhou Li, Lingyu Wang, Yu Qiu, Yangyao Chen, and Ke Xu contributed to the overall interpretation of the results and various aspects of the analysis.

\end{addendum}
\clearpage

\noindent{\bf References}
\bigskip
\bigskip

\bibliography{main}{}
\bibliographystyle{plainnat}

\clearpage

\section*{\centering \large Supplementary material}
\setcounter{figure}{0}

\begin{Extended Data Figure}
	\centering
	\includegraphics[width=\linewidth]{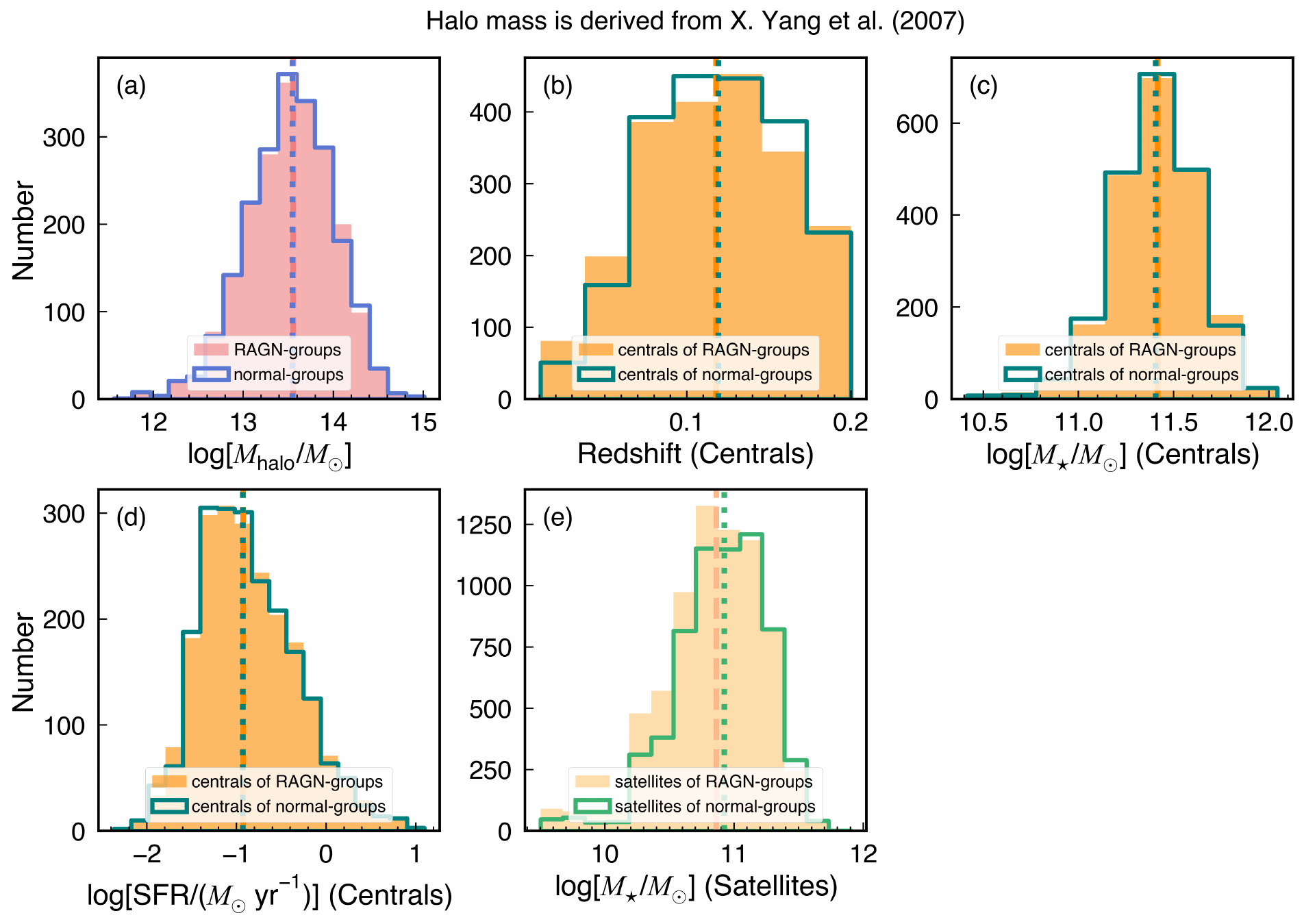}
 \caption{Distribution of host halo mass (panel (a)), redshift of central galaxies (panel (b)), stellar mass of central galaxies (panel (c)), star formation rate of central galaxies (panel (d)), 
 and stellar mass of satellite galaxies (panel (e)) in RAGN-groups (solid histogram) and normal-groups (empty histogram) with halo mass derived from Ref.\cite{Yang2007}. Red/blue distributions present the properties of the entire galaxy group, orange/green distributions mean the properties of central galaxies, light orange/light green distributions describe the properties of satellite galaxies.
 The dashed and dotted lines represent the median value of distributions of RAGN-groups and normal-groups, respectively.
 }
 \label{efig:histYangRAGN}
\end{Extended Data Figure}

\begin{Extended Data Figure}
	\centering
	\includegraphics[width=\linewidth]{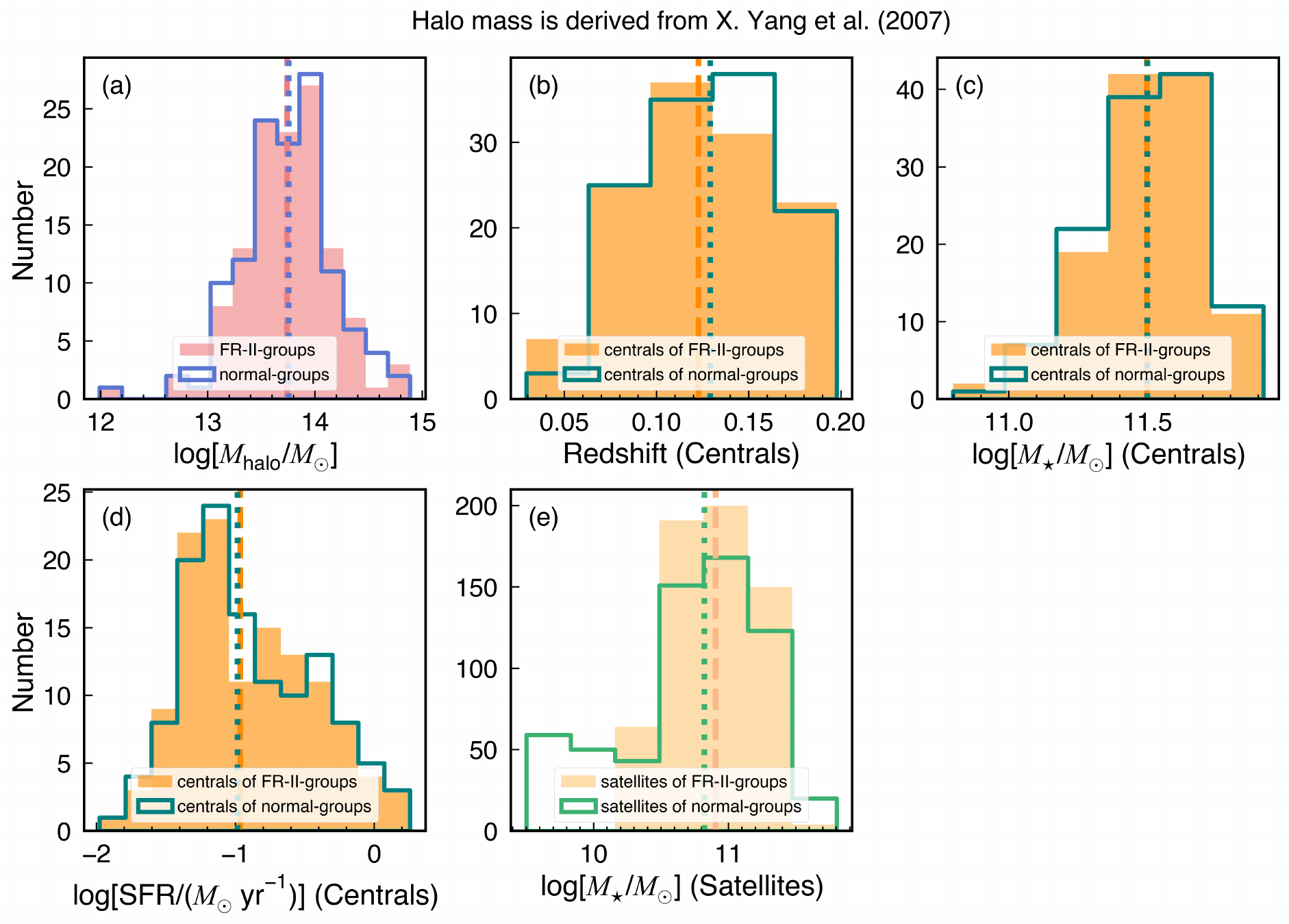}
 \caption{Same as Extended Data Figure \ref{efig:histYangRAGN} but show the comparison between FR-II-groups and normal-groups.}
 \label{efig:histYangFRII}
\end{Extended Data Figure}

\begin{Extended Data Figure}
	\centering
	\includegraphics[width=\linewidth]{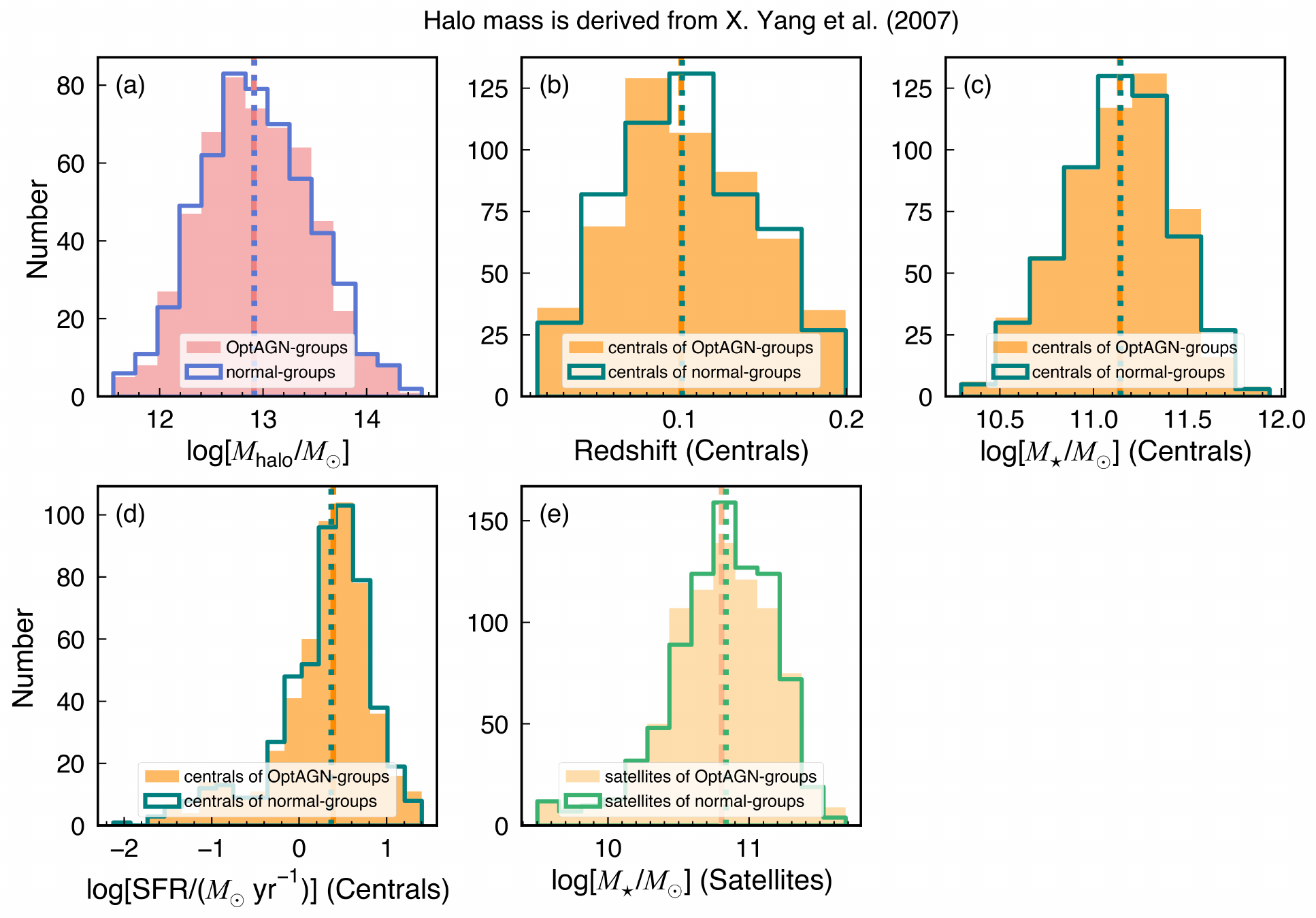}
 \caption{Same as Extended Data Figure \ref{efig:histYangRAGN} but show the comparison between OptAGN-groups and normal-groups.}
 \label{efig:histYangBPT}
\end{Extended Data Figure}

\begin{Extended Data Figure}
	\centering
	\includegraphics[width=\linewidth]{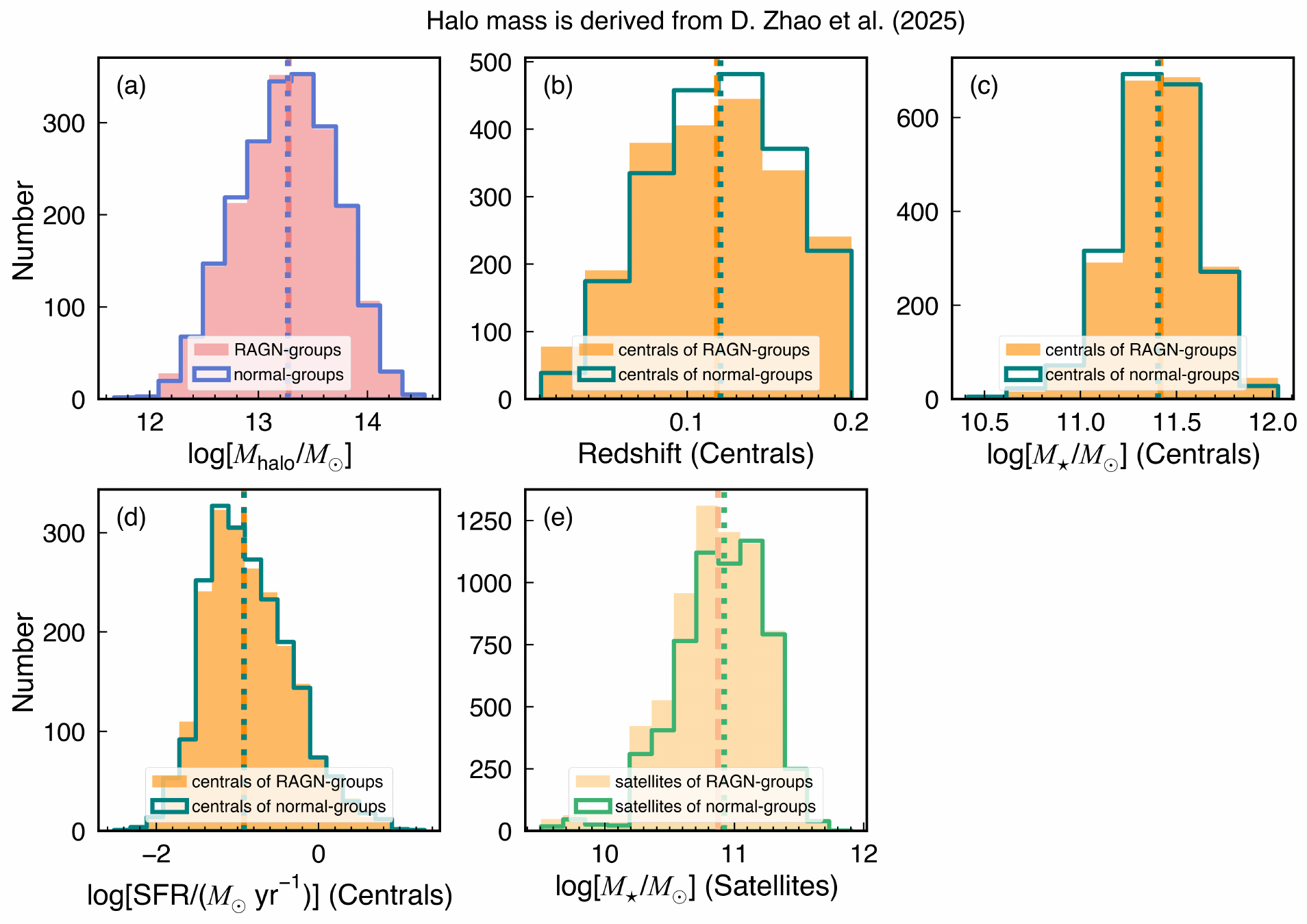}
 \caption{Same as Extended Data Figure \ref{efig:histYangRAGN} but with halo mass derived from Ref.\cite{ZhaoDY2025}.}
 \label{efig:histZhaoRAGN}
\end{Extended Data Figure}

\begin{Extended Data Figure}
	\centering
	\includegraphics[width=\linewidth]{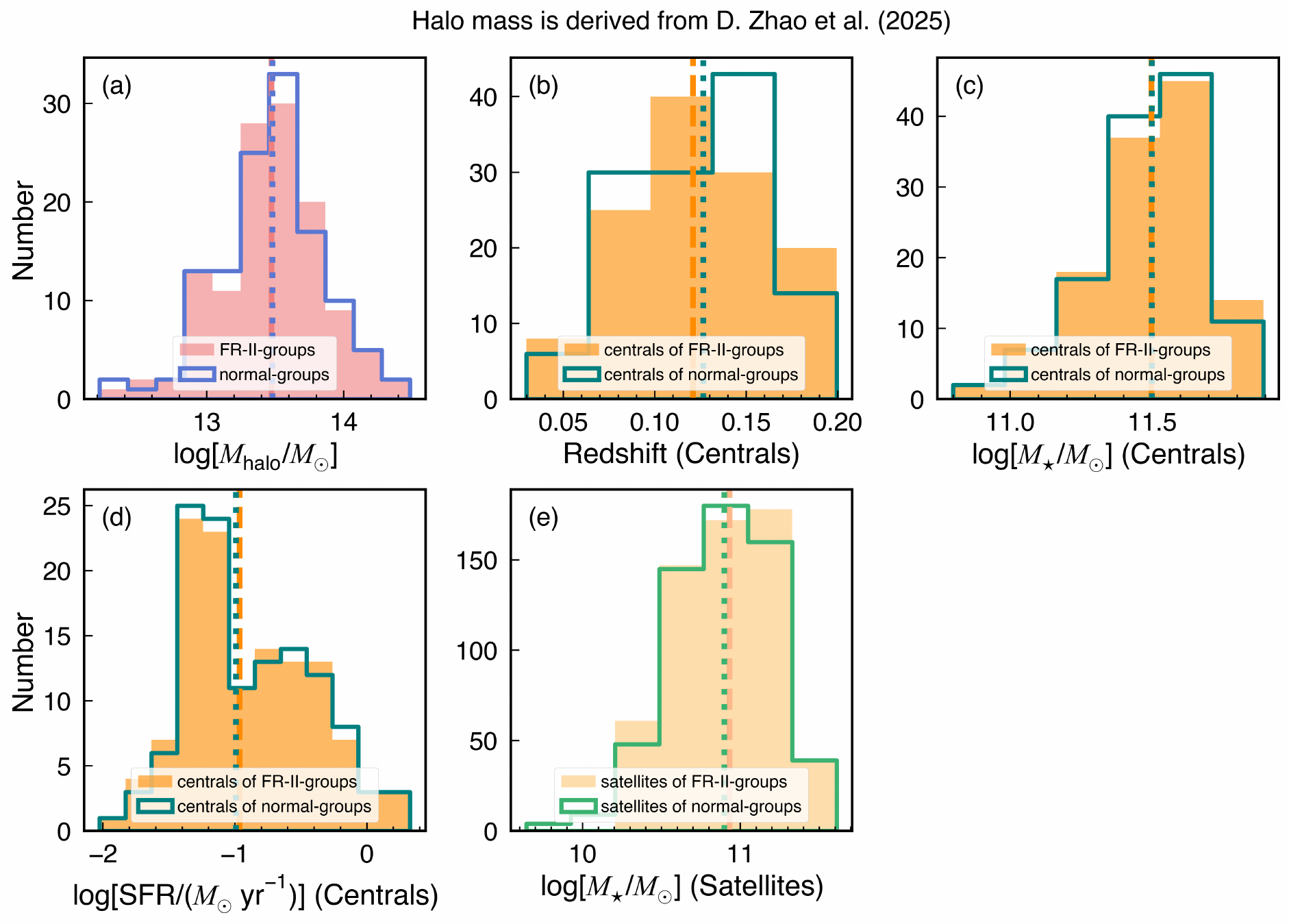}
 \caption{Same as Extended Data Figure \ref{efig:histZhaoRAGN} but show the comparison between FR-II-groups and normal-groups.}
 \label{efig:histZhaoFRII}
\end{Extended Data Figure}

\begin{Extended Data Figure}
	\centering
	\includegraphics[width=\linewidth]{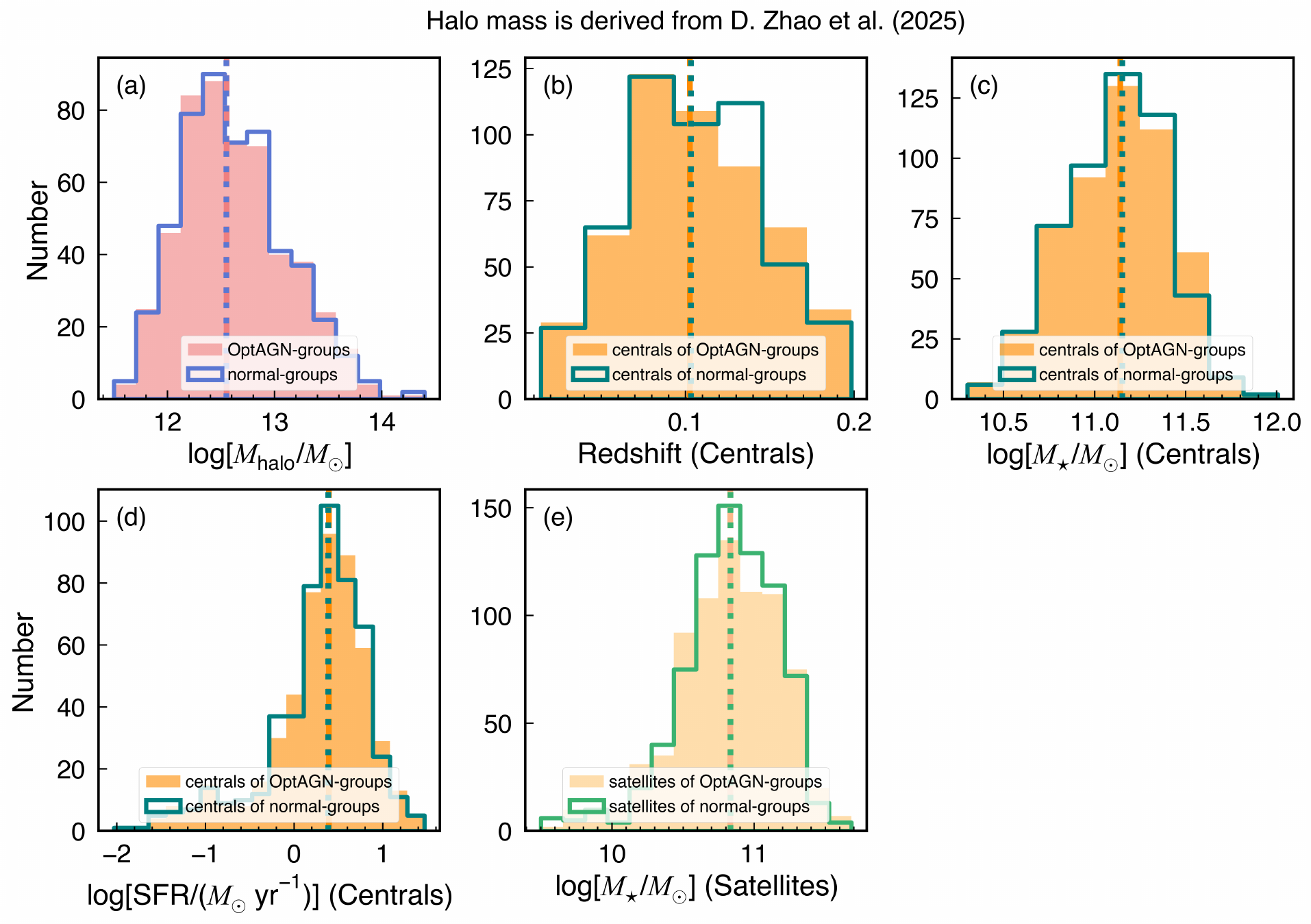}
 \caption{Same as Extended Data Figure \ref{efig:histZhaoRAGN} but show the comparison between OptAGN-groups and normal-groups.}
 \label{efig:histZhaoBPT}
\end{Extended Data Figure}

\begin{Extended Data Figure}
	\centering
\includegraphics[width=\linewidth]{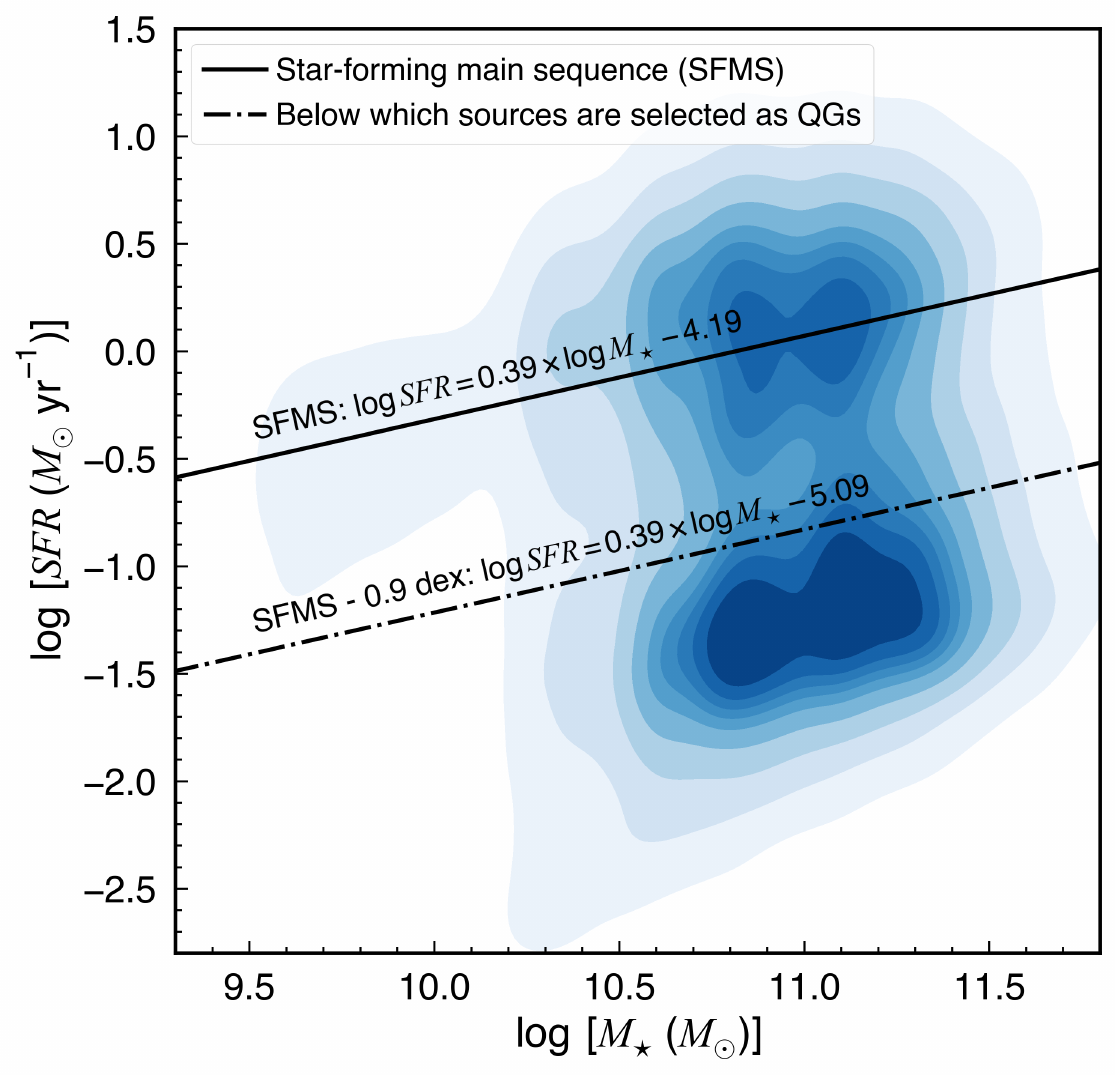}
 \caption{Star formation rate (\textit{SFR}) versus stellar mass ($M_\star$) for all mass-complete galaxies in the SDSS galaxy groups at $0.01 \leq z \leq 0.20$ (blue contour). The black solid line represents the star-forming main sequence (SFMS), the black dash-dot line is 0.9 dex lower than the SFMS, below which sources are selected as quiescent galaxies (QGs). 
 }
	\label{efig1}
\end{Extended Data Figure}

\end{document}